\documentclass[twocolumn,preprintnumbers,amsmath,amssymb,showpacs]{revtex4}

\usepackage{graphicx}
\usepackage{dcolumn}
\usepackage{bm}

\newcommand{\remark}[1]{{\tt[#1]}}

\newcommand{\whatref}[1]{ \remark{\mbox{$\backslash$ref\{?\}}}}

\newcommand{\refeq}[1]{(\ref{#1})}

\newcommand{\basinAv}{\beta}
\newcommand{\initial}{i}

\begin{document}

\title{Random close packing revisited: How many ways can we pack 
frictionless disks?}

\author{Ning Xu$^1$}
\author{Jerzy Blawzdziewicz$^1$} 
\author{Corey S. O'Hern$^{1,2}$}
\affiliation{$^1$~Department of Mechanical Engineering, Yale University, 
New Haven, CT 06520-8284.\\
$^2$~Department of Physics, Yale University, New Haven, CT 06520-8120.\\
}
\date{\today}

\begin{abstract}
We create collectively jammed (CJ) packings of $50$-$50$ bidisperse
mixtures of smooth disks in 2d using an algorithm in which we
successively compress or expand soft particles and minimize the total
energy at each step until the particles are just at contact.  We focus
on small systems in 2d and thus are able to find nearly all of the
collectively jammed states at each system size.  We decompose the
probability $P(\phi)$ for obtaining a collectively jammed state at a
particular packing fraction $\phi$ into two composite functions: 1)
the density of CJ packing fractions $\rho(\phi)$, which only depends
on geometry and 2) the frequency distribution $\basinAv(\phi)$, which
depends on the particular algorithm used to create them.  We find that
the function $\rho(\phi)$ is sharply peaked and that $\basinAv(\phi)$
depends exponentially on $\phi$.  We predict that in the infinite
system-size limit the behavior of $P(\phi)$ in these systems is
controlled by the density of CJ packing fractions---not the frequency
distribution.  These results suggest that the location of the peak in
$P(\phi)$ when $N \rightarrow \infty$ can be used as a
protocol-independent definition of random close packing.
\end{abstract}

\maketitle

\section{Introduction}
\label{introduction}

Developing a statistical mechanical description of dense granular
materials, structural and colloidal glasses, and other jammed systems
\cite{book} composed of discrete macroscopic grains is a difficult,
long-standing problem.  These amorphous systems possess an enormously
large number of possible jammed configurations; however, it is not
known with what probabilities these configurations occur since these
systems are not in thermal equilibrium.  The possible jammed
configurations do not occur with equal probability---in fact, some are
extremely rare and others are highly probable.  Moreover, the
likelihood that a given jammed configuration occurs depends on the
protocol that was used to generate it.

Despite difficult theoretical challenges, there have been a number of
experimental and computational studies that have investigated
jammed configurations in a variety of systems. The experiments include
studies of static packings of ball bearings \cite{bernal,scott},
slowly shaken granular materials \cite{knight,phillippe}, sedimenting
colloidal suspensions \cite{kegel}, and compressed colloidal glasses
\cite{zhu}.  The numerical studies include early Monte Carlo
simulations of dense liquids \cite{finney}, collision dynamics of
growing hard spheres \cite{lubachevsky}, serial deposition of granular
materials under gravity \cite{pavlovitch,tkachenko,barker}, various
geometrical algorithms \cite{jodrey,clarke,speedy}, compression
and expansion of soft particles followed by energy minimization
\cite{ohern_long}, and other relaxation methods \cite{zinchenko}.

\begin{figure}
\scalebox{0.5}{\includegraphics{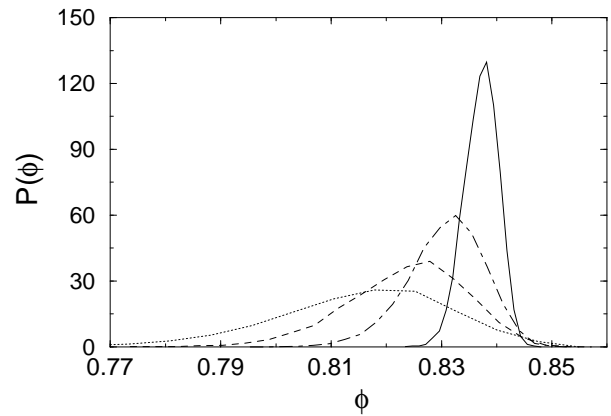}}%
\vspace{-0.18in}
\caption{\label{fig:TD_largeN} The probability distribution $P(\phi)$
to obtain a collectively jammed state at packing fraction $\phi$ in 2d
bidisperse systems with $N=18$ (dotted), $32$ (dashed), $64$
(dot-dashed), and $256$ (solid).}
\vspace{-0.22in}
\end{figure}

The early experimental and computational studies found that dense
amorphous packings of smooth, hard particles frequently possess
packing fractions near random close packing $\phi_{\rm rcp}$, which is
approximately $0.64$ in $3d$ monodisperse systems \cite{berryman} and
$0.84$ in the $2d$ bidisperse systems discussed in this work
\cite{speedy,ohern_short}.  However, more recent studies have
emphasized that the packing fraction attained in jammed systems can
depend on the process used to create them.  Different protocols select
particular configurations from a distribution of jammed states with
varying degrees of positional and orientational order
\cite{torquato_2000}.

Recent studies of hard particle systems have also shown that different
classes of jammed states exist with different properties \cite{ts}.
For example, in {\it locally} jammed (LJ) states, each particle is
unable to move provided all other particles are held fixed; however,
groups of particles can still move collectively.  In contrast, in
collectively jammed (CJ) states neither single particles nor groups of
particles are free to move (excluding `floater' particles that do not
have any contacts).  Thus, CJ states are more `jammed' than LJ states.

In this article we focus exclusively on the properties of collectively
jammed states.  These states are created using an energy minimization
procedure \cite{ohern_long,ohern_short} for systems composed of particles
that interact via soft, finite-range, purely repulsive, and spherically
symmetric potentials.  Energy minimization is combined with
successive compressions and decompressions of the system to find states
that cannot be further compressed without producing an overlap of the
particles.  As explained in Sec.\ \ref{methods}, this procedure yields
collectively jammed states of the equivalent hard-particle system.

In previous studies of collectively jammed states created using the
energy minimization method, we showed that the probability
distribution of collectively jammed packing fractions narrows as the
system size increases and becomes a $\delta$-function located at
$\phi_0$ in the infinite system-size limit
\cite{ohern_long,ohern_short}.  We found that $\phi_0$ was similar to
values quoted previously for random close packing \cite{berryman}. The
narrowing of the distribution of CJ packing fractions as the system
size increases is shown in Fig.~\ref{fig:TD_largeN} for 2d bidisperse
systems.  However, it is still not clear why this happens.  Why it is
so difficult to obtain a collectively jammed state with $\phi \ne
\phi_0$ in the large system limit?  One possibility is that very few
collectively jammed states exist with $\phi \ne \phi_0$.  Another
possibility is that collectively jammed states do exist over a range
of packing fractions, but only those with packing fractions near
$\phi_0$ are very highly probable.

Below, we will address this question and other related problems by
studying the distributions of collectively jammed states in small
bidisperse systems in 2d.  For such systems we we will be able to
generate nearly all of the collectively jammed states.  Enumeration of
nearly all CJ states will allow us to decompose the probability
density $P(\phi)$ to obtain a collectively jammed state at a
particular packing fraction $\phi$ into two  contributions
\begin{equation}
\label{factorization of probability density}
P(\phi)=\rho(\phi)\basinAv(\phi).
\end{equation}
The factor $\rho(\phi)$ in the above equation represents the density
of collectively jammed states (i.e., $\rho(\phi)d \phi$ measures how
many distinct collectively jammed states exist within in a small range
of packing fractions $d \phi$).  The factor $\basinAv$ denotes the
effective frequency (i.e., the counts averaged over a small region of
$\phi$) with which these states occur.

We note that the density of states $\rho(\phi)$ is determined solely
by the topological features of configurational space; it is thus
independent of the the protocol used to generate these states.  In
contrast, the quantity $\basinAv(\phi)$ is protocol dependent, because
it records the average frequency with which a CJ state at $\phi$
occurs for a given protocol.  For example, for algorithms that allow
partial thermal equilibration during compression and expansion, the
frequency distributions are shifted to larger $\phi$ compared to
those that do not involve such equilibration.  

The decomposition \refeq{factorization of probability density} will
allow us to determine which contribution, $\rho(\phi)$ or
$\basinAv(\phi)$, controls the shape of the probability distribution
$P(\phi)$ in the large system limit.  Others have studied the inherent
structures of hard-sphere liquids and glasses, but have not addressed
this specific question \cite{speedy2,bowles}.  We will show below that
$\rho(\phi)$ controls the width of the distribution of CJ states in
the infinite system-size limit.  We also have some evidence that the
location of the peak in $P(\phi)$ in the large $N$ limit is also
determined by the large $N$ behavior of $\rho(\phi)$.  We will also
argue that for many procedures the protocol-dependence of the
frequency distribution $\basinAv(\phi)$ is too weak to substantially
shift the peak in $P(\phi)$ for large systems.  Thus, our results
suggest that for a large class of algorithms the location of the peak
in $P(\phi)$ can be used as a protocol-independent definition of
random close packing in the infinite system size limit.

\begin{figure}
\scalebox{0.4}{\includegraphics{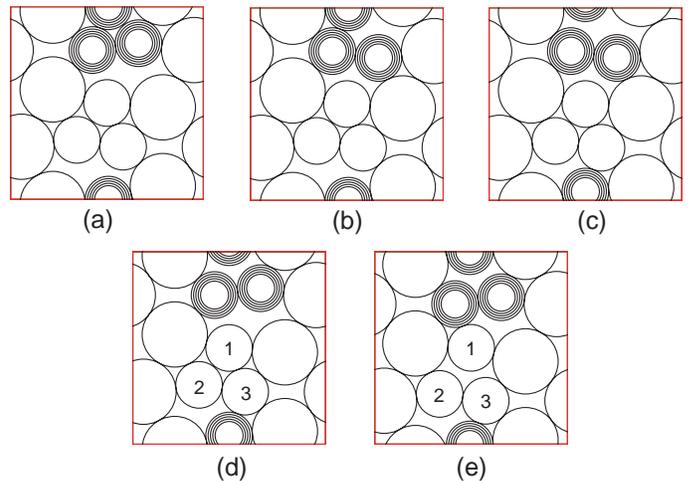}}%
\vspace{-0.18in}
\caption{\label{fig:phi} Five distinct collectively jammed states that
exist at the same packing fraction $\phi=0.81073$ for a 2d bidisperse
system with $N=12$ particles.  In (a) through (d), unshaded particles
are in the same positions, while the shaded particles are in different
locations from panel to panel.  Particles labeled $1$, $2$, and $3$
are in different positions in panels (d) and (e), while the other
particles are in the same positions.}
\vspace{-0.22in}
\end{figure}

\section{Methods}
\label{methods}

Our goal is to enumerate the collectively jammed configurations in 2d
bidisperse systems composed of smooth, repulsive disks.  We will focus
on bidisperse mixtures composed of $N/2$ large and $N/2$ small
particles with a diameter ratio $\eta=1.4$ because it has been shown that
these systems do not easily crystallize or phase separate
\cite{speedy,ohern_long}.  We consider system sizes in the range $N=4$
to $256$ particles.  For $N \le 10$, we were able to find nearly all
of the collectively jammed states.  For $N = 12$ ($14$) we found more 
than $90\%$ ($60\%$) of the total number.  Since the number of
collectively jammed states grows so rapidly with $N$, we are not able
to calculate a large fraction of the CJ states for $N > 14$, but as we
will show below, we can still make strong conclusions about the shape
of the distribution of CJ states in large systems.

We utilize an energy-minimization procedure to create collectively
jammed states \cite{ohern_long}.  We assume that the particles
interact via the purely repulsive linear spring potential
\begin{equation}
\label{spring potential}
V(r_{ij}) =\frac{\epsilon}{2}(1-r_{ij}/d_{ij})^2\Theta(d_{ij}/r_{ij}-1),
\end{equation}
where $\epsilon$ is the characteristic energy scale, $r_{ij}$ is the
separation of particles $i$ and $j$, $d_{ij}=\left(d_i + d_j \right)
/2$ is their average diameter, and $\Theta(x)$ is the Heaviside step
function.  The potential \refeq{spring potential}, is nonzero only for
$r_{ij} < d_{ij}$, i.e., when the particles overlap.  Jammed states
are obtained by successively growing or shrinking particles followed
by relaxation via potential energy minimization until all particles
(excluding floaters) in the system are just at contact.  In these
prior studies, we showed that the distribution of collectively jammed
states does not depend sensitively on the shape of the repulsive
potential $V(r_{ij})$.  Note that our process for creating jammed
states differs from the {\it fixed} volume energy minimization
procedure implemented in Ref.~\cite{ohern_long}.  In the description
below, the energies and lengths are measured in units of $\epsilon$
and the diameter of the smaller particle $d_1$.

\begin{figure}
\scalebox{0.5}{\includegraphics{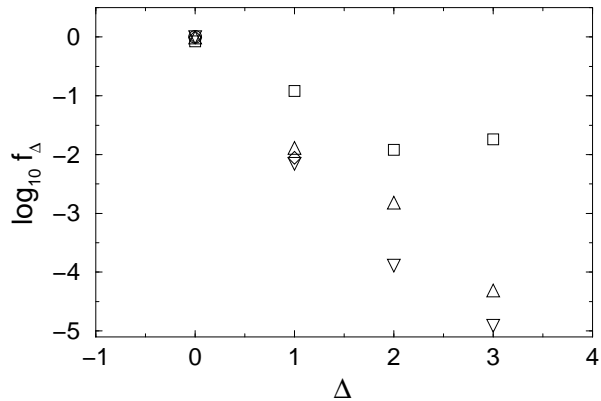}}%
\vspace{-0.18in}
\caption{\label{fig:fraction} Fraction $f_{\Delta}$ of distinct
collectively jammed states with an excess number of contacts $\Delta =
N_c - N_c^{\rm min}$ over a range of system sizes, $N=6$ (circles), $8$
(squares), $10$ (diamonds), $12$ (upward triangles), and $14$
(downward triangles).  For $N=6$, all CJ states have $\Delta=0$.}
\vspace{-0.22in}
\end{figure}

For each independent trial, the procedure begins by choosing a random
configuration of $N$ particles at an initial packing fraction
$\phi_\initial$ in a square box with unit length and periodic boundary
conditions.  The positions of the centers of the particles are
uncorrelated and distributed uniformly in the box.  We have found that
the results do not depend on the initial volume fraction
$\phi_\initial$ as long it is significantly below the peak in
$\rho(\phi)$. We chose $\phi_\initial=0.60$ for most system sizes.

After initializing the systems, we find the nearest local potential
energy minimum using the conjugate gradient algorithm \cite{numrec}.
We terminate the energy minimization procedure when either of the
following two conditions is satisfied: 1) two successive conjugate
gradient steps $n$ and $n+1$ yield nearly the same total potential
energy per particle, $(V_{n+1} - V_{n})/V_n < \delta = 10^{-16}$ or 2)
the total potential energy per particle is extremely small, $V_{n+1} <
V_{\rm min} = 10^{-16}$.

Following the potential energy minimization, we decide whether the
system should be compressed or expanded to find the jamming threshold.
If $V_{n+1} > V_{\rm max} = 2 \times 10^{-16}$, particles have nonzero
overlap and thus small and large particles are reduced in size by
$\Delta d_1 = d_1 \Delta \phi /(2 \phi)$ and $\Delta d_2 = \eta \Delta
d_1$, respectively.  If, on the other hand, $V_{n+1} \le V_{\rm min}$, the
system is below the jamming threshold and all particles are thus
increased in size.  After the system has been expanded or compressed,
it is relaxed using potential energy minimization and the process is
repeated.  Each time the procedure switches from expansion to
contraction or vice versa, the packing fraction increment $\Delta
\phi$ is reduced by a factor of $2$.  The initial expansion rate was
$\Delta \phi_\initial = 10^{-4}$. 

When the total potential energy per particle falls within the range
$V_{\rm max} > V > V_{\rm min}$, the process is terminated and the
`jammed' packing fraction is recorded. If the final state contains
floater particles with $2$ or fewer contacts, we remove them, minimize
the total potential energy, and slightly compress or expand the
remaining particles to find the jamming threshold.  Note that the
final configurations are slightly compressed with overlaps in the
range $10^{-9} < 1- r_{ij}/d_{ij} < 10^{-8}$.  We have verified that
our results do not depend strongly on the parameters $V_{\rm min}$,
$V_{\rm max}$, and $\Delta \phi_\initial$.

For each system-size $N$, this process is repeated using $n_t'$
independent random initial conditions and the resulting jammed
configurations are analyzed to determine whether they are collectively
jammed and unique.

\begin{figure}
\scalebox{0.5}{\includegraphics{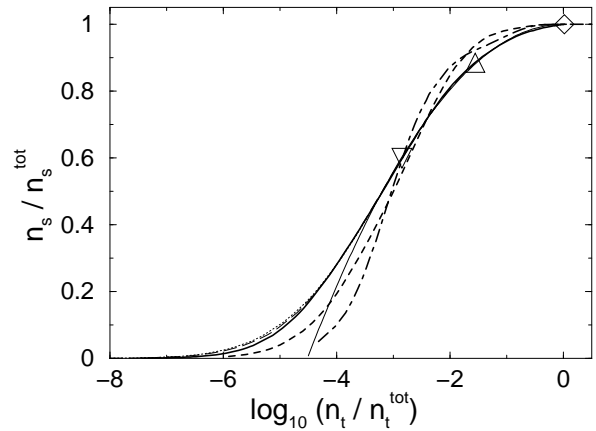}}%
\vspace{-0.18in}
\caption{\label{fig:Ns_Nt} Fraction of CJ states $n_s/n_s^{\rm tot}$
versus the ratio of the number of trials $n_t$ to the total number of
trials $n_t^{\rm tot}$ required to find all CJ states $n_s^{\rm tot}$ for
several system sizes: $N=6$ (dot-dashed), $8$ (dashed), $10$ (thick
solid), $12$ (long dashed), and $14$ (dotted).  The curves for $N=10$,
$12$, and $14$ collapse. The thin solid line is a least-squares fit to
Eq.~\refeq{ns}.  The diamond, upward triangle, and downward triangle
symbols give the maximum number of trials attempted for $N=10$, $12$,
and $14$, respectively.}
\vspace{-0.22in}
\end{figure}

\section{Analysis of Jammed States}
\label{analysis}

To verify if a given final configuration is collectively jammed we
analyze the eigenvalue spectra of the dynamical (or rigidity) matrix
\cite{ohern_long}, $M_{i\alpha,j\beta}$, where the indices $i$ and $j$
refer to the particles and $\alpha,\beta=x,y$ represent the Cartesian
coordinates.  For a system with $N_f$ floaters and $N'=N-N_f$
particles forming a connected network the indices $i$ and $j$ range
from $1$ to $N'$.  Thus, the dynamical matrix has $dN'$ rows and
columns, where $d=2$ is the spatial dimension.  By differentiating the
interparticle potential we find that the elements of the dynamical
matrix with $i \ne j$ are given by \cite{tanguy}
\begin{equation}
\label{offdiagonal}
M_{i\alpha,j\beta} = -\frac{t_{ij}}{r_{ij}}\left( \delta_{\alpha \beta} 
- {\hat r}_{ij\alpha} {\hat r}_{ij\beta} \right) - c_{ij} {\hat r}_{ij\alpha} 
{\hat r}_{ij\beta}, 
\end{equation}
where $t_{ij} = \partial V/\partial r_{ij}$ and $c_{ij} = \partial^2
V/ \partial r_{ij}^2$, while those with $i=j$ are given by
\begin{equation}
\label{diagonal}
M_{i\alpha,i\beta} = - \sum_j M_{i\alpha,j\beta}.
\end{equation}

The dynamical matrix \refeq{offdiagonal} and \refeq{diagonal} has
$dN'$ real eigenvalues $\{ \xi_i \}$, $d$ of which are zero due to
translational invariance of the system.  In a collectively jammed
state no set of particle displacements is possible without creating an
overlapping configuration; therefore the dynamical matrix has exactly
$dN'-d$ nonzero eigenvalues.  In our simulations we use the criterion
$|\xi_i| > \xi_{\rm min}$ for nonzero eigenvalues, where $\xi_{\rm
min} = 10^{-6}$ is the noise threshold for our eigenvalue
calculations.

We note that our energy minimization algorithm for creating jammed
states does occasionally yield a configuration that is not
collectively jammed. These states, however, are not considered in the
current study.  The number of trials that yield collectively jammed
states out of the original $n_t'$ trials is denoted $n_t$.  The
fraction of trials that give locally but not collectively jammed
states $(n_t'-n_t)/n_t'$ decreases with increasing system size from
$\approx 5\%$ at $N=6$ to less than $1\%$ for $N>12$.

\begin{figure}
\scalebox{0.5}{\includegraphics{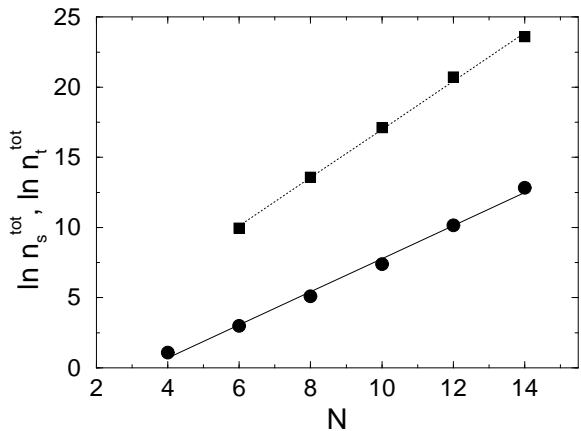}}%
\vspace{-0.18in}
\caption{\label{fig:Ns_N} The total number of distinct CJ states $n_s^{\rm
tot}$ (circles) and the number of trials required to find them (squares) 
versus system-size $N$.  The solid and dotted lines have slopes equal
to $1.2$ and $1.7$, respectively.}
\vspace{-0.22in}
\end{figure}

We determine whether two collectively jammed states are distinct by
comparing the sorted lists of the nonzero eigenvalues of their
respective dynamical matrices.  If the relative difference between two
corresponding eigenvalues differs by more than $\xi_{\rm
diff}=10^{-3}$, the configurations are treated as distinct.  By
comparing the topology of the network of particle contacts in
a representative sample of CJ states, we have found that this
criterion is sufficient to reliably determine whether two states are
distinct or identical.

This procedure allows us to determine the number $n_s$ of distinct
collectively jammed states at each fixed number of independent trials
$n_t$.  As expected, if two CJ states have different packing
fractions, they are distinct, with different contact networks and
dynamical modes.  This property holds with very high numerical
precision---the packing-fraction difference of $10^{-13}$ already
assures that the two states are distinct.

However, it is not true that all collectively jammed states with the
same packing fraction are identical.  For example, the CJ states shown
in Fig.~\ref{fig:phi} have the same packing fraction, but they possess
different contact networks and eigenvalue spectra.  This is a clear
demonstration that two collectively jammed configurations at the same
packing fraction can have very different structural properties.

We have also calculated the total number of contacts between particles
$N_c$, i.e. the number of bonds that satisfy $r_{ij} < d_{ij}$, in our
slightly compressed jammed configurations.  We find that the number of
contacts in the collectively jammed states satisfies the relation
\cite{zinchenko,donev_2005}
\begin{equation}
\label{condition}
N_c \ge N_c^{\rm min} = 2 \left(dN' - d + 1\right).
\end{equation}
The minimum number of contacts required for mechanical stability of
the system $N_c^{\rm min}$ can be calculated by equating the number of
degrees of freedom to the number of constraints.  Note that an extra
constraint is required to prevent particle expansion.  We have found
that nearly all of the collectively jammed states have $N_c = N_c^{\rm
min}$; fewer than $1\%$ of these states have $N_c > N_c^{\rm min}$ as
shown in Fig.~\ref{fig:fraction}.  All configurations that are not
collectively jammed have fewer contacts than $N_c^{\rm min}$.

\begin{table}
\caption{\label{tab:table1} Maximum number of trials performed $n_t^{\rm max}$ 
and fraction of CJ states obtained $(n_s/n_s^{\rm tot})_{\rm max}$ versus 
system size $N$.}
\begin{tabular}{cccc}
$N$ & & $n_t^{\rm max}$ & $(n_s/n_s^{\rm tot})_{\rm max}$\\
\hline
$6$ & & $10^6$ & $1.0$ \\
$8$ & & $10^6$ & $1.0$ \\
$10$ & \hspace{0.1in} & $29 \times 10^6$&  $1.0$  \\
$12$ & & $28 \times 10^6$&  $0.90$  \\
$14$ & & $26 \times 10^6$&  $0.60$   \\
\end{tabular}
\end{table}
  
\section{Results}
\label{results}

In the preceding two sections, we described our methods for generating
and counting distinct collectively jammed states.  We will now present
the results from these analyses.  We will first discuss how the number
of CJ states depends on parameters such as the number of trials and
system size.  We then decompose the probability density of obtaining a
CJ state at a given packing fraction (Eq.~\refeq{factorization of
probability density}) into the density $\rho(\phi)$ of CJ packing
fractions and their frequency distribution $\basinAv(\phi)$.  We also
consider under what conditions all of the possible CJ states can be
enumerated and determine whether strong conclusions can be made about
the distributions of CJ states in large systems even though complete
enumeration is not possible.

\begin{figure}
\scalebox{0.5}{\includegraphics{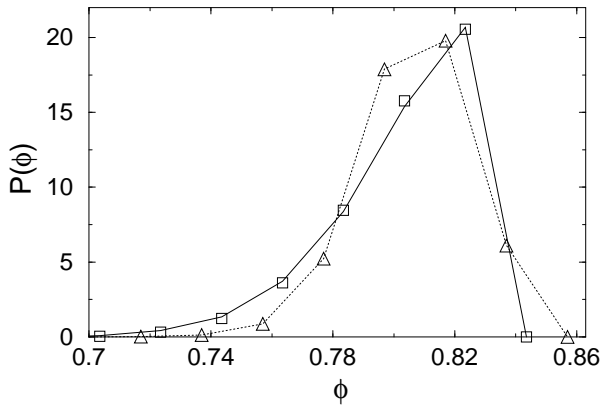}}%
\vspace{-0.18in}
\caption{\label{fig:TD_N} Probability distribution $P(\phi)$ for
obtaining a CJ state at $\phi$ for $N=10$ (solid line) and $N=14$
(dotted line) at $(n_s/n_s^{\rm tot})_{\rm max}$.  The
distributions at $n_s/n_s^{\rm tot} = 0.05$ (squares for $N=10$ and
triangles for $N=14$) overlap those with larger $n_s/n_s^{\rm tot}$.}
\vspace{-0.22in}
\end{figure}

Our studies of the number of distinct CJ states $n_s$ versus the
number of independent trials $n_t$ led to several surprising
observations.  First, we find that these systems possess a significant
fraction of rare CJ states and thus an exponentially large number of
trials are required to obtain nearly all states.  Second, a master
curve appears to describe $n_s(n_t)$ for systems with $N \ge 10$, as
shown in Fig.~\ref{fig:Ns_Nt}.  (Each data point in this figure was
obtained by averaging over at least $100$ distinct permutations of the
$n_t$ trials.)  Our numerical results indicate that when $n_s$ is more
than about $20\%$ of the total number of distinct CJ states $n_s^{\rm
tot}$, the curve $n_s(n_t)$ can be accurately approximated by
\begin{equation}
\label{ns}
\frac{n_s}{n_s^{\rm tot}} = 1 - A \left[ \log_{10} \left( \frac{n_t}
{n_t^{\rm tot}} \right) \right]^2,
\end{equation}
where $A \approx 0.05$.  

Our direct computations for small systems ($N=6$, $8$, and $10$) and
numerical fits to the master curve \refeq{ns} for $N=12$ and $14$
indicate that both $n_t^{\rm tot}$ and $n_s^{\rm tot}$ increase
exponentially with system size as shown in Fig.~\ref{fig:Ns_N}.
However, both these quantities remain finite for any finite system. In
particular, for the smallest system sizes, we increased the total
number of trials by at least a factor of $10$ and did not find any new
collectively jammed states.  We also used several different algorithms
for generating CJ states, e.g. compression and expansion of particles
followed by relaxation using molecular dynamics with dissipative
forces along ${\hat r}_{ij}$ and frictional forces perpendicular to
${\hat r}_{ij}$, and these did not lead to any new CJ states that were
not already found using the protocol described in Sec.~\ref{methods}.
The maximum number of trials and fraction of CJ states obtained are
provided in Table~\ref{tab:table1}.  

\begin{figure}
\scalebox{0.5}{\includegraphics{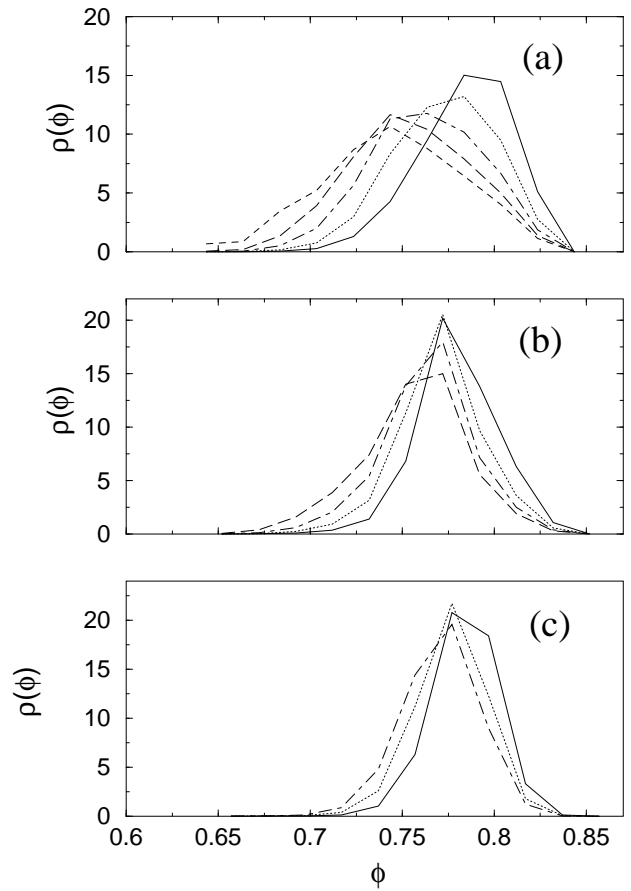}}%
\vspace{-0.18in}
\caption{\label{fig:DOS} Density of collectively jammed packing
fractions $\rho(\phi)$ for (a) $N=10$, (b) $12$, and (c) $14$ at
$n_s/n_s^{\rm tot} = 0.2$ (solid), $0.4$ (dotted), $0.6$ (dot-dashed),
$0.8$ (long-dashed), and $1.0$ (dashed).}
\vspace{-0.22in}
\end{figure}

As indicated in Eq.\ \refeq{factorization of probability density}, the
probability distribution $P(\phi)$ for obtaining a collectively jammed
state at a particular packing fraction $\phi$ can be factorized into
two composite functions: the density of CJ states $\rho(\phi)$ and the
frequency $\basinAv(\phi)$ with which these states occur.  In our
simulations, the distribution $P(\phi)$ is calculated from the
relation
\begin{equation}
\label{probability}
P(\phi) d \phi = \frac{n_P(\phi + d\phi) - n_P(\phi)}{n_t}.
\end{equation}
Here $n_P(\phi)$ is the total number of CJ states (counting all
repetitions of the same state) with packing fractions below $\phi$.
The density of CJ states is evaluated using an analogous relation
\begin{equation}
\label{densit_of_states}
\rho(\phi) d\phi = \frac{n_s(\phi + d \phi) - n_s(\phi)}{n_s},
\end{equation}
where $n_s(\phi)$ is the number of distinct CJ states that have been
detected in the packing-fraction range below $\phi$. (In fact, we have
used the number of distinct packing fractions to define $\rho(\phi)$
in place of the number of distinct CJ states $n_s(\phi)$. However,
this does not affect our results because distinct states with the
same $\phi$ are rare in 2d bidisperse systems.)  We note that both the
probability density \refeq{probability} and the density of CJ states
\refeq{densit_of_states} are normalized to 1.  The frequency
distribution $\basinAv(\phi) = P(\phi)/\rho(\phi)$ is normalized
accordingly.

Below, we show how $P(\phi)$, $\rho(\phi)$, and $\basinAv(\phi)$
depend on the fraction of CJ states $n_s/n_s^{\rm tot}$ and system
size $N$.  To plot these distributions, we used $10$ bins with the
endpoint of the final bin located at the largest CJ packing fraction
$\phi_{\rm max}$ for each $N$.  We recall that the distribution of CJ
packing fractions $\rho(\phi)$ does not depend on the protocol used to
generate the CJ states.  The protocol dependence of the distribution
$P(\phi)$ is captured by the frequency distribution $\basinAv(\phi)$.

The probability distribution $P(\phi)$ of CJ states is shown in
Fig.~\ref{fig:TD_N} for two small systems $N=10$ and $14$.  The
results indicate that $P(\phi)$ depends very weakly on the fraction
$n_s/n_s^{\rm tot}$ of CJ states obtained---only $5\%$ of the CJ
states are required to capture accurately the shape of $P(\phi)$ for
these systems.  This result holds for all system sizes we studied,
which implies that the distribution of CJ states can be measured
reliably even in large systems \cite{ohern_long,ohern_short}.  Note
that the width and the location of the peak in $P(\phi)$ do not change
markedly over the narrow range of $N$ shown in Fig.~\ref{fig:TD_N}.

To see significant changes in $P(\phi)$, the system size must be
varied over a larger range.  $P(\phi)$ for $N=18$, $32$, $64$, and
$256$ are shown in Fig.~\ref{fig:TD_largeN} at fixed number of trials
$n_t=10^4$.  The width of the distribution narrows and the peak
position shifts to larger $\phi$ as the system size increases.  In
Ref.~\cite{ohern_long}, we found that $P(\phi)$ for this 2d bidisperse
system becomes a $\delta$-function located at $\phi_0 = 0.842$ in the
infinite-system-size limit.  What causes $P(\phi)$ to narrow to a
$\delta$-function located at $\phi_0$ when $N \rightarrow \infty$?  Is
the shape of the distribution $P(\phi)$ determined primarily by the
density of states $\rho(\phi)$, or does the frequency distribution
$\basinAv(\phi)$ play a significant role in determining the
width and location of the peak?  We will shed light on these
questions below.

\begin{figure}
\scalebox{0.5}{\includegraphics{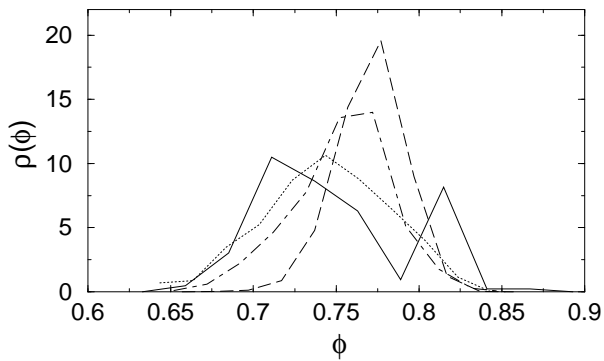}}%
\vspace{-0.18in}
\caption{\label{fig:compare_DOS} Density of collectively jammed packing
fractions $\rho(\phi)$ for $N=8$ (solid), $10$ (dotted), $12$ (dot-dashed), 
and $14$ (long-dashed) at $(n_s/n_s^{tot})_{\rm max}$.}
\vspace{-0.22in}
\end{figure}

We first show results for $\rho(\phi)$ and $\basinAv(\phi)$ as
functions of the fraction $n_s/n_s^{\rm tot}$ of distinct CJ states
obtained.  In Fig.~\ref{fig:DOS}, $\rho(\phi)$ is shown for several
small systems.  In contrast to the total distribution $P(\phi)$, the
density of states $\rho(\phi)$ depends on $n_s/n_s^{\rm tot}$
significantly.  For $N=10$, a system for which we can calculate nearly
all of the CJ states, the curve $\rho(\phi)$ reaches its final height
and width when $n_s/n_s^{\rm tot} \approx 0.5$.  However, its shape
still slowly evolves as $n_s/n_s^{\rm tot}$ increases above $0.5$;
the low-$\phi$ part of the curve increases while the high-$\phi$ side
decreases.  This implies that the {\it rare} CJ states are not
uniformly distributed in $\phi$, but are more likely to occur at low
packing fractions below the peak in $\rho(\phi)$.  Similar results for
$\rho(\phi)$ as functions of $n_s/n_s^{\rm tot}$ are found for $N=12$
and $14$. By comparing $\rho(\phi)$ at fixed $n_s/n_s^{tot}$, we also
find that $\rho(\phi)$ narrows with increasing $N$.  To further
demonstrate that $\rho(\phi)$ narrows, the density of states is
plotted in Fig.~\ref{fig:compare_DOS} for several system sizes at
$(n_s/n_s^{\rm tot})_{\rm max}$ listed in Table \ref{tab:table1}.

\begin{figure}
\scalebox{0.5}{\includegraphics{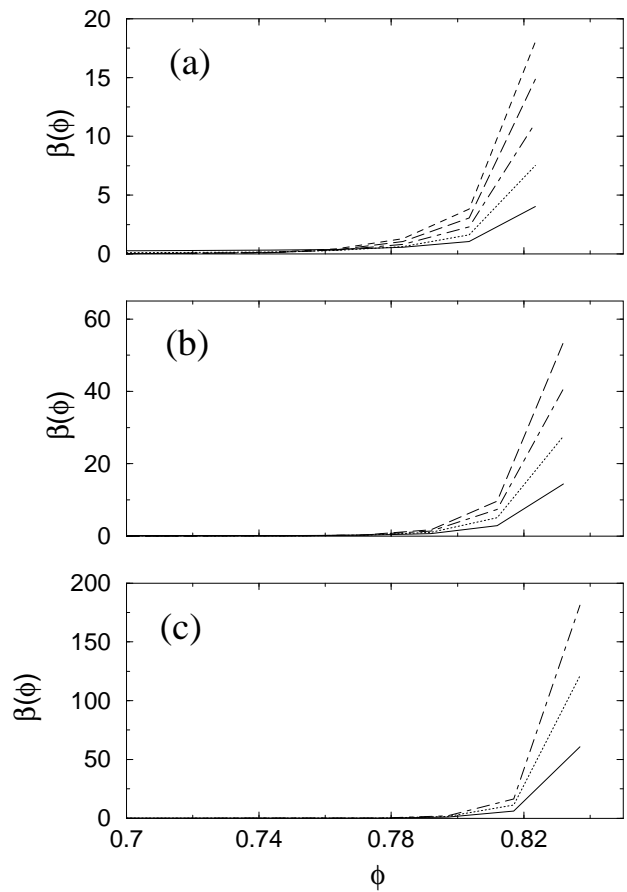}}%
\vspace{-0.1in}
\caption{\label{fig:BOA} Frequency distribution $\basinAv(\phi)$ for 
(a) $N=10$, (b) $12$, and (c) $14$ for $n_s/n_s^{\rm tot} = 0.2$ (solid), 
$0.4$ (dotted), $0.6$ (dot-dashed), $0.8$ (long-dashed), and $1.0$ (dashed).}
\vspace{-0.22in}
\end{figure}

The dependence of the frequency distribution $\basinAv(\phi)$ on the
system size $N$ and the fraction $n_s/n_s^{\rm tot}$ of CJ states
obtained is illustrated in Fig.~\ref{fig:BOA}.  The results show that
in contrast to the functions $P(\phi)$ and $\rho(\phi)$, the
distribution $\basinAv(\phi)$ achieves its maximal value at the
highest packing fraction for which CJ states exist $\phi_{\rm max}$.
By comparing $\basinAv(\phi)$ for different system sizes at fixed
$n_s/n_s^{\rm tot}$ we find that $\phi_{max}$ increases with
increasing $N$.

The frequency distribution $\basinAv(\phi)$ becomes more strongly
peaked at $\phi_{\max}$ as $n_s/n_s^{\rm tot}$ increases.  The
evolution of $\basinAv(\phi)$ with $n_s/n_s^{\rm tot}$ can be
explained by noting that $\basinAv(\phi) \equiv P(\phi)/\rho(\phi)$
and that $P(\phi)$ does not depend on $n_s/n_s^{\rm tot}$ for
$n_s/n_s^{\rm tot}\gtrsim 0.05$ according to the results shown in
Fig.\ \ref{fig:TD_N}.  The density of states $\rho(\phi)$ and the
frequency distribution $\basinAv(\phi)$ must therefore behave in
opposite ways to maintain constant $P(\phi)$.  As shown earlier in
Fig.~\ref{fig:DOS}, the peak in $\rho(\phi)$ widens (for $n_s/n_s^{\rm
tot} < 0.5$) and shifts to lower packing fractions as $n_s/n_s^{\rm
tot}$ increases.  Thus, the distribution $\basinAv(\phi)$ must
decrease at low packing fractions and build up at large packing
fractions with increasing $n_s/n_s^{\rm tot}$.

\begin{figure}
\scalebox{0.5}{\includegraphics{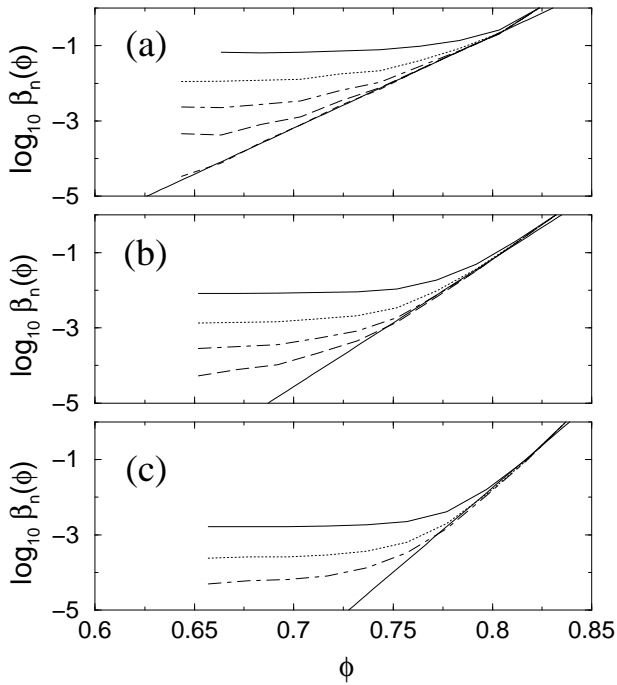}}%
\vspace{-0.18in}
\caption{\label{fig:BOA_compare} Frequency distribution $\basinAv_n(\phi)$
normalized by the peak value for (a) $N=10$, (b) $N=12$, and (c) $N=14$
at $n_s/n_s^{\rm tot} = 0.2$ (solid), $0.4$ (dotted), $0.6$ (dot-dashed),
$0.8$ (long-dashed), and $1.0$ (dashed).  Least squares fits to exponential 
curves (thin solid lines) are also shown for the largest $n_s/n_s^{\rm tot}$
at each $N$.}
\vspace{-0.22in}
\end{figure}

In Fig.~\ref{fig:BOA_compare}, we show the frequency distribution
$\basinAv_n(\phi)=\basinAv(\phi)/\basinAv_{\rm max}$, which is
normalized by the peak value $\basinAv_{\rm max}$.  The results are
plotted on a logarithmic scale.  The frequency distribution varies
strongly with $\phi$; CJ states with small packing fractions are rare
and those with large packing fractions $(\phi \approx 0.83)$ occur
frequently. We find that $\basinAv_n(\phi)$ is exponential over an
expanding range of $\phi$ as $n_s/n_s^{\rm tot}$ increases.  For
$N=10$, $\basinAv_n(\phi)$ increases exponentially over nearly the
entire range of $\phi$ at $n_s/n_s^{\rm tot}=1$.  We see similar
behavior for $N=12$ and $14$ in panels (b) and (c) of
Fig.~\ref{fig:BOA_compare}; thus we expect $\basinAv_n(\phi)$ to be
exponential as $N_s/N_s^{\rm tot} \rightarrow 1$ for $N>10$.  We have
calculated least-squares fits to
\begin{equation}
\label{exponential fits of beta}
\basinAv_n = A_{\beta}
\exp (B_{\beta} \phi)
\end{equation}
for the largest $n_s/n_s^{tot}$ at each system size. As pointed out
above, the frequency distribution becomes steeper with increasing $N$;
we find that $B_{\beta}$ increases by a factor of $3.5$ as $N$
increases from $10$ to $18$ (not shown).  Note that reasonable
estimates of $B_{\beta}$ can be obtained even at fairly low values of
$n_s/n_s^{\rm tot}$.

\begin{figure}
\scalebox{0.5}{\includegraphics{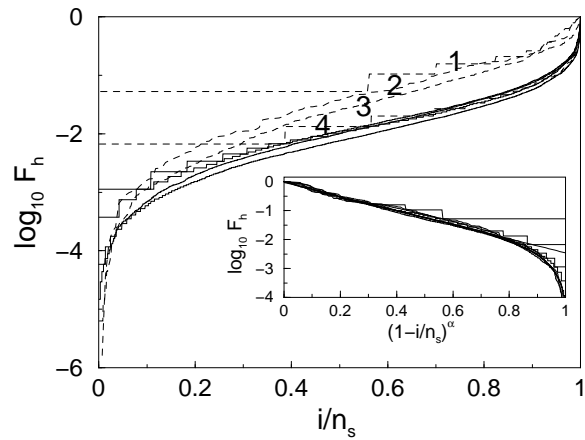}}%
\vspace{-0.18in}
\caption{\label{fig:newhit} Cumulative probability distribution $F_h$
of CJ states in a narrow range of packing fractions $d \phi$ for N=12.
The index $i$ denotes the position of the state in a list ordered by
the frequency of occurrence and $n_s$ is the total number
of states in the given interval.  The solid lines correspond to bins
centered on $\phi=0.73$, $0.75$, $0.77$, and $0.79$; the dashed lines
labeled $1$, $2$, $3$, and $4$ correspond to bins centered on
$\phi=0.65$, $0.81$, $0.83$, and $0.69$, respectively.  The width of
each bin is $\Delta \phi = 0.02$.  The inset shows that the data are
well-described by Eq.\ (\protect\ref{fit to cumulative distribution}),
where $A_F \approx 2.4$ and $\alpha$ varies from $0.3$ to $0.4$.}
\vspace{-0.22in}
\end{figure}

We showed in Fig.~\ref{fig:BOA_compare} that the frequency
distribution is not uniform in $\phi$; in contrast, it increases
exponentially with $\phi$.  Fig.~\ref{fig:newhit} shows another
striking result; the frequency distribution is also highly nonuniform
within a narrow range of $\phi$.  In this figure, we plot the
cumulative distribution $F_h$ of the probabilities of jammed states
in an narrow interval $d \phi$ versus the index $i$ in a list of
all distinct states in $d \phi$ ordered by the value of the
probability of each state.  The data for several different intervals
appear to collapse onto a stretched exponential form,
\begin{equation}
\label{fit to cumulative distribution}
F_h = \exp \left[ -A_F
\left(1-i/n_s\right)^{\alpha}\right],
\end{equation}
where $n_s$ is the number of distinct CJ states within $d \phi$
and the exponent $\alpha$ varies from $0.3$ to $0.4$.  These results
clearly demonstrate that CJ states can occur with very different
frequencies even if they have similar packing fractions.

\begin{figure}
\scalebox{0.5}{\includegraphics{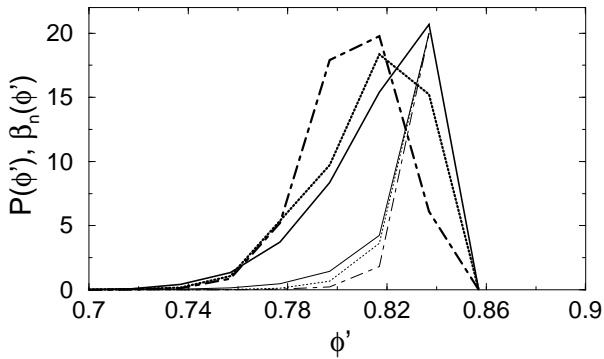}}%
\vspace{-0.18in}
\caption{\label{fig:compare_TD} $P(\phi')$ (thick lines) and
$\basinAv_n(\phi')$ (thin lines) for $N=10$ (solid), $12$ (dotted), and
$N=14$ (dot-dashed) at $(n_s/n_s^{tot})_{\rm max}$, where $\phi' =
\phi + \Delta \phi$.  $P(\phi)$ and $\basinAv_n(\phi)$ for $N=10$, $12$,
and $14$ have been shifted by $\Delta \phi=0.013$, $0.005$, and $0$
respectively, so that $\basinAv_{\rm max}$ for the three system sizes
coincide. $\basinAv_n(\phi)$ for each $N$ has also been amplified by a
factor of $\approx 20$.}
\vspace{-0.in}
\end{figure}

From our studies of small systems, we find that both the density
$\rho(\phi)$ of CJ packing fractions and the frequency distribution
$\basinAv(\phi)$ narrow and shift to larger packing fractions as the
system size increases.  (See Figs.~\ref{fig:DOS} and
~\ref{fig:BOA_compare}.)  How do these changes in $\rho(\phi)$ and
$\basinAv(\phi)$ affect the total distribution $P(\phi)$ and can we
determine which changes dominate in the large system limit?  To shed
some light on these questions, we consider the position of the peak in
$P(\phi)$ with respect to the maximal packing fraction of CJ states
$\phi_{\rm max}$ for several system sizes.  In the absence of
changes in $\rho(\phi)$ as a function of $\phi-\phi_{\rm max}$, the
maximum of $P(\phi)$ should shift {\it toward\/} $\phi=\phi_{\rm \max}$
with increasing system size, because the frequency distribution
$\basinAv$ becomes more sharply peaked at $\phi_{\rm max}$ according
to the results in Fig.\ \ref{fig:BOA_compare}.  However, as shown in
Fig.~\ref{fig:compare_TD}, we find the opposite behavior over the range
of system sizes we considered: the peak of $P(\phi)$ shifts {\it
away\/} from $\phi_{\max}$.  This suggests that the density of states,
not the frequency distribution, plays a larger role in determining the
location of the peak in $P(\phi)$ in these systems.

\begin{figure}
\scalebox{0.5}{\includegraphics{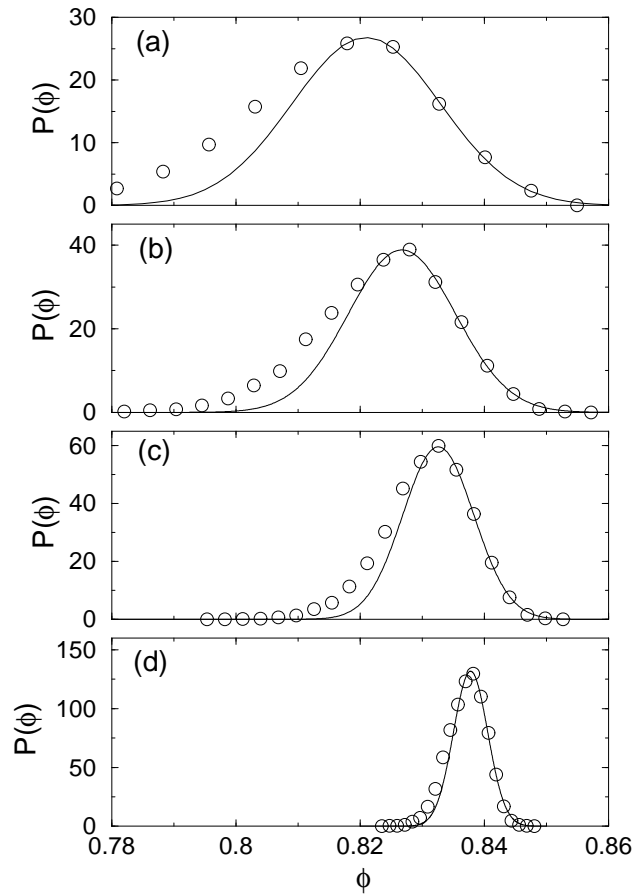}}%
\vspace{-0.18in}
\caption{\label{fig:Gaussian_N} The distribution $P(\phi)$ of CJ
states for (a) $N=18$, (b) $32$, (c) $64$, and (d) $256$ are depicted
using circles.  The solid lines are least squares fits of the
large-$\phi$ side of $P(\phi)$ to Gaussian distributions.}
\vspace{-0.1in}
\end{figure}

Additional conclusions about the relative roles of the the density of
states and the frequency distribution on the position and width of
$P(\phi)$ can be drawn from our observation that the frequency
distribution $\basinAv$ is an exponential function of $\phi$ (c.f.\
the discussion of results in Fig.\ \ref{fig:BOA_compare}) and that
$P(\phi)$ is Gaussian for sufficiently large systems (as shown in
\cite{ohern_long} and illustrated in Fig. \ref{fig:Gaussian_N}).  If
we assume that the exponential form of the frequency distribution
\refeq{exponential fits of beta} remains valid in the
large-system-limit, the density of states
$\rho(\phi)=P(\phi)/\basinAv(\phi)$ is also Gaussian with the
identical width $\sigma(N)$.
The location of the peak in $P(\phi)$ is 
\begin{equation}
\label{position of peak in P}
\phi^*_P(N)=\phi^*_\rho(N)+B_{\beta}(N)\sigma(N),
\end{equation}
where $\phi_\rho^*(N)$ is the location of the peak in $\rho(\phi)$.
In previous studies \cite{ohern_long}, we found that the width of
$P(\phi)$ scaled as $\sigma \sim N^{-\Omega}$, with $\Omega \approx
0.55$.  We have also some indication that $B_{\beta} \sigma^2$
decreases with increasing system size: $B_{\beta} \sigma^2 = 0.017$ at
$N=12$ compared to $0.012$ at $N=18$.  However, we are not currently
able to estimate $B_{\beta}$ in the large $N$ limit.

If the system-size dependence of $B_{\beta}$ is weaker than $N^{2
\Omega}$, the quantity $B_{\beta} \sigma^2$ will tend to zero and the
frequency distribution will not influence the location of the peak in
$P(\phi)$.  In this case $P(\phi)$ becomes independent of the
frequency distribution in the limit $N\to\infty$ for a class of
protocols that are characterized by a similar frequency distribution
$\basinAv$ as our present protocol.  Thus, as our preliminary results
suggest, random close packing can be defined as the location of the
peak in $\rho(\phi)$ when $N \rightarrow \infty$, and this definition
is completely independent of on the algorithm used to generate the CJ
states.  In the opposite case, where the system-size dependence of
$B_{\beta}$ is stronger than $N^{2 \Omega}$, the position of the peak
in $P(\phi)$ results from a subtle interplay between the density of
states and the frequency distribution.  However, even in this case one
can argue that the dependence of the position of the peak only weakly
depends on the protocol: a shift of the peak position requires an
exponential change in the frequency distribution $\basinAv(\phi)$.

\section{Conclusions}
\label{conclusions}

We have studied the possible collectively jammed configurations that
occur in small 2d periodic systems composed of smooth purely repulsive
bidisperse disks.  The CJ states were created by successively
compressing or expanding soft particles and minimizing the total
energy at each step until the particles were just at contact.  By
studying small 2d systems, we were able to enumerate nearly all of the
collectively jammed states at each system size and therefore decompose
the probability distribution $P(\phi)$ for obtaining a CJ state at a
particular packing fraction $\phi$ into the density $\rho(\phi)$ of CJ
packing fractions and their frequency distribution $\basinAv(\phi)$.
The distribution $\basinAv(\phi)$ depends on the particular protocol used
to generate the CJ configurations, while $\rho(\phi)$ does not.  This
decomposition allowed us to study how the protocol-independent
$\rho(\phi)$ and protocol-dependent $\basinAv(\phi)$ influence the shape
of $P(\phi)$.

These studies yielded many important and novel results.  First, the
probability distribution $P(\phi)$ of CJ states is nearly independent
of $n_s/n_s^{\rm tot}$, and thus it can be measured reliably even in large
systems.  This finding validates several previous measurements of
$P(\phi)$ \cite{ohern_long,ohern_short}.  Second, the number of
distinct CJ states grows exponentially with system size.  In addition,
a large fraction of these configurations are extremely rare and thus
an exponentially large number of trials are required to find all of
the CJ states.  Third, the frequency distribution $\basinAv(\phi)$ are
nonuniform and increase exponentially with $\phi$.  We also found that
even over a narrow range of $\phi$, the frequency with which
particular CJ states occur is strongly nonuniform and involves a large
number of exponentially rare states.  Finally, we have shown that
$P(\phi)$ becomes Gaussian in the large $N$ limit.  Since $P(\phi) =
\rho(\phi) \basinAv(\phi)$ and $\basinAv(\phi)$ is exponential, we expect
that $\rho(\phi)$ is also Gaussian and controls the width of
$P(\phi)$ for large $N$. We also have preliminary results
that suggest that the contribution from $\basinAv(\phi)$ to the shift of
the peak in $P(\phi)$ decreases with increasing $N$.  We expect that
$\rho(\phi)$ will determine the location of the peak in $P(\phi)$ in
the large $N$ limit, and thus it is a robust protocol-independent
definition of random close packing in this system.

\section{Future Directions}
\label{future}

Several interesting questions have arisen from this work that will be
addressed in our future studies.  First, we have shown that the
frequency with which CJ states occur is highly nonuniform.  It is
important to ask whether the rare states can be neglected in
analyses of static and dynamic properties of jammed and nearly jammed
systems.  For example, we have shown that $P(\phi)$ is insensitive to
the fraction $n_s/n_s^{\rm tot}$ of CJ states obtained and thus
$P(\phi)$ is not influenced by the rare CJ states.  However, rare CJ
states may be important in determining the {\it dynamical} properties
of jammed and glassy systems if these states are associated with
`passages' or `channels' from one frequently occurring state to another.
Moreover, an analysis of the density of states and the frequency
distribution (both as a function of $\phi$ and locally in $\phi)$ may
shed light on the phase space evolution of glassy systems during
the aging process.

A closely related question is what topological or geometrical features
of configurational phase space give rise to the exponentially
varying frequency distribution?  Can one, for example, uniquely assign
a specific volume in configurational space to each jammed state?  A
candidate for such a quantity is the volume $\Omega_c$ of configuration 
space in which each point is connected by a continuous path without particle 
overlap to a particular CJ state.  It is likely that those CJ
states with large $\Omega_c$ will occur frequently for a typical
compaction algorithm, while those with small $\Omega_c$ will be rare.
 
\begin{figure}
\scalebox{0.5}{\includegraphics{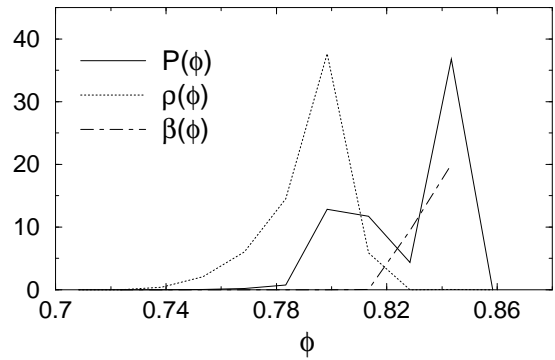}}%
\vspace{-0.18in}
\caption{\label{fig:monoN24} $P(\phi)$ (solid), $\rho(\phi)$ (dotted),
and $\basinAv(\phi)$ (dot-dashed) for a 2d monodisperse system with $N=24$.}
\vspace{-0.22in}
\end{figure}

Another important question is whether the results for $\rho(\phi)$,
$\basinAv(\phi)$, and $P(\phi)$ found in 2d bidisperse systems also hold
for other systems such as {\it monodisperse} systems in 2d and 3d.
Does $\rho(\phi)$ still control the behavior of $P(\phi)$ or does the
frequency distribution play a more dominant role in determining
$P(\phi)$?  To begin to address these questions, we have enumerated
nearly all of the distinct collectively jammed states and calculated
$P(\phi)$, $\rho(\phi)$, and $\beta(\phi)$ in small 2d periodic cells
containing $N=4$ to $32$ equal-sized particles.

In our preliminary studies, we have found several significant
differences between 2d monodisperse and bidisperse systems, which
largely stem from the fact that partially ordered states occur
frequently in the monodisperse systems.  First, in 2d monodisperse
systems there is an abundance of distinct CJ states that exist at the
same packing fraction.  For example, in a monodisperse systems with
$N=24$, multiple distinct states occur at $19\%$ of the CJ packing
fractions compared to less than $1\%$ in bidisperse systems with
$N=14$.  Second, for the system sizes studied, quantitative features
of the distributions of CJ states depend on whether $N$ is even or
odd.  Third, $P(\phi)$ can possess {\it two} strong peaks.  For
example, two peaks in $P(\phi)$ occur at $\phi_1 \approx 0.805$ and
$\phi_2 \approx 0.844$ for $N=24$ as shown in Fig.~\ref{fig:monoN24}.
Moreover, the large-$\phi$ peak that corresponds to partially ordered
configurations is a factor of three taller than the small-$\phi$ peak
that corresponds to amorphous configurations.  Finally, the maximum in
$\basinAv(\phi)$ coincides with the large-$\phi$ peak in $P(\phi)$ and
$\basinAv(\phi)$ decays very rapidly as $\phi$ decreases.  As shown in
Fig.~\ref{fig:monoN24}, the rapid decay of $\basinAv(\phi)$ significantly
suppresses the contribution of the peak in $\rho(\phi)$ to the total
distribution $P(\phi)$.  Thus, $\basinAv(\phi)$, which depends on the
protocol used to generate the CJ states, may strongly influence the
total distribution $P(\phi)$ even in moderately sized 2d monodisperse
systems.

Many open questions concerning monodisperse systems in 2d will be
answered in a forthcoming article \cite{ning}.  We will measure the
shape of $P(\phi)$ as a function of system size and predict whether
$\rho(\phi)$ or $\basinAv(\phi)$ controls the width and location of the
peak (or peaks) in the large $N$ limit.  The fact that $\basinAv(\phi)$
strongly influences $P(\phi)$ at small and moderate system sizes
explains why it has been so difficult to determine random close
packing in 2d monodisperse systems \cite{berryman}---different
protocols have yielded different values for $\phi_{\rm rcp}$
\cite{torquato_2000,donev}.  \\

\subparagraph{Acknowledgments}

Financial support from NSF grant numbers CTS-0348175 (JB) and
DMR-0448838 (NX,CSO) is gratefully acknowledged.  We also thank Yale's
High Performance Computing Center for generous amounts of computer
time.

\end{document}